\documentclass[]{spie}  %>>> use for US letter paper
\usepackage{amsmath,amsfonts,amssymb}
\usepackage{graphicx}
\usepackage[colorlinks=true, allcolors=blue]{hyperref}

\begin{document}

\title{MITS: the Multi-Imaging Transient Spectrograph for SOXS}

\author[a]{Adam Rubin}
\author[b]{Sagi Ben-Ami}
\author[a]{Ofir Hershko}
\author[a]{Michael Rappaport}
\author[a]{Oz Diner}
\author[a]{Avishay Gal-Yam}
\author[c]{Sergio Campana}
\author[d]{Riccardo Claudi}
\author[e]{Pietro Schipani}
\author[c]{Matteo Aliverti}
\author[d]{Andrea Baruffolo}
\author[d]{Federico Biondi}
\author[f]{Anna Brucalassi}
\author[e]{Giulio Capasso}
\author[g,h]{Rosario Cosentino}
\author[i]{Francesco D'Alessio}
\author[c]{Paolo D'Avanzo}
\author[j,k]{Hanindyo Kuncarayakti}
\author[h]{Matteo Munari}
\author[h]{Salvatore Scuderi}
\author[i]{Fabrizio Vitali}
\author[l]{Jani Achr\'en}
\author[m]{Jos\'e Antonio Araiza-Duran}
\author[n]{Iair Arcavi}
\author[c]{Andrea Bianco}
\author[d]{Enrico Cappellaro}
\author[e]{Mirko Colapietro}
\author[e]{Massimo Della Valle}
\author[e]{Sergio D'Orsi}
\author[d]{Daniela Fantinel}
\author[o]{Johan Fynbo}
\author[c]{Matteo Genoni}
\author[p]{Mika Hirvonen}
\author[j]{Jari Kotilainen}
\author[k]{Tarun Kumar}
\author[c]{Marco Landoni}
\author[p]{Jussi Lehti}
\author[q]{Gianluca Li Causi}
\author[d]{Luca Marafatto}
\author[k]{Seppo Mattila}
\author[c]{Giorgio Pariani}
\author[m]{Giuliano Pignata}
\author[d]{Davide Ricci}
\author[c]{Marco Riva}
\author[d]{Bernardo Salasnich}
\author[h]{Ricardo Zanmar-Sanchez}
\author[r]{Stephen Smartt}
\author[d]{Massimo Turatto}
\affil[a]{Weizmann Institute of Science, Rehovot, Israel}
\affil[b]{Harvard-Smithsonian Center for Astrophysics, Cambridge, USA}
\affil[c]{INAF - Osservatorio Astronomico di Brera, Merate, Italy}
\affil[d]{INAF - Osservatorio Astronomico di Padova, Padua, Italy}
\affil[e]{INAF - Osservatorio Astronomico di Capodimonte, Naples, Italy}
\affil[f]{European Southern Observatory, Garching, Germany}
\affil[g]{INAF -- Fundaci\'{o}n Galileo Galilei, Bre\~{n}a Baja, Spain}
\affil[h]{INAF - Osservatorio Astrofisico di Catania, Catania, Italy}
\affil[i]{INAF - Osservatorio Astronomico di Roma, Rome, Italy}
\affil[j]{FINCA - Finnish Centre for Astronomy with ESO, Turku, Finland}
\affil[k]{University of Turku, Turku, Finland}
\affil[l]{Incident Angle Oy, Turku, Finland}
\affil[m]{Universidad Andres Bello, Santiago, Chile}
\affil[n]{Tel Aviv University, Tel Aviv, Israel}
\affil[o]{Dark Cosmology Centre, Copenhagen, Denmark}
\affil[p]{ASRO - Aboa Space Research Oy, Turku, Finland}
\affil[q]{INAF - Istituto di Astrofisica e Planetologia Spaziali, Rome, Italy}
\affil[r]{Queen's University Belfast, Belfast, UK}

\authorinfo{Corresponding author A.R., adam.rubin@weizmann.ac.il}

\maketitle

\begin{abstract}
The Son Of X-Shooter (SOXS) \cite{schipani_new_2016} is a medium resolution spectrograph $(R\sim4500)$ proposed for the ESO 3.6 m NTT. We present the optical design of the UV-VIS arm of SOXS which employs high efficiency ion-etched gratings used in first order $(m=1)$ as the main dispersers. The spectral band is split into four channels which are directed to individual gratings, and imaged simultaneously by a single three-element catadioptric camera. The expected throughput of our design is $>60\%$ including contingency. The SOXS collaboration expects first light in early 2021. This paper is one of several papers presented in these proceedings\cite{soxsschipani,soxsaliverti,soxsbiondi,soxsbrucalassi,soxscapasso,soxsclaudi,soxsricci,soxssanchez,soxsvitali} describing the full SOXS instrument.
\end{abstract}

\section{Introduction}

The Son Of X-Shooter (SOXS) \cite{schipani_new_2016} is a medium resolution spectrograph $(R\sim4500)$\footnote{For design purposes we assume a $1''$ slit. A $1''$ is also assumed for the quoted resolution.} proposed for the ESO 3.6 m NTT. Originally both arms of SOXS were based on the X-Shooter 4-C design \cite{delabre_astronomical_1989}. The 4-C design uses a cross-dispersed \'echelle grating to achieve medium resolution at higher orders (10-20). In 2016 the Weizmann Institute of Science joined SOXS with a proposal for redesigning the VIS arm of the spectrograph to make use of high efficiency gratings working at order $m=1$. Here we present this concept and design, which passed a preliminary design review (PDR) in July 2017, and will be presented at a final design review (FDR) in the summer of 2018.

The main science goal of SOXS is the follow-up and study of astronomical transients of all types. The flood of photometrically identified transients expected to be supplied by upcoming surveys (e.g. ZTF, ATLAS, PanSTARRS, LSST) will require dedicated spectroscopic facilities such as SOXS. The X-Shooter observational characteristics are highly desirable, as evidenced by X-Shooter being one of the most oversubscribed ESO instruments.

To serve these scientific goals the following requirement were defined. Here we show those requirements that are relevant to the VIS arm design:

\begin{itemize}
	
	\item The UV-VIS arm will cover $350-850$ nm, with an overlap between the VIS and NIR arms that spans 50 nm ($800-850$ nm).
	\item Limiting (Vega) magnitudes for stellar objects, long slit: Considering $t_{exp}$=1 hr, S/N=10, $1''$ slit, $0.8''$ seeing, no moon and airmass 1.2: V=20 and J=19. 
	\item The peak efficiency shall be $>25\%$ and the efficiency at any wavelength shall be $>8\%$ except in the dichroic crossover regimes.
	\item Spectral Resolution: the slit width – resolution product shall be RS $>3500$ over the whole spectral range. 
	 
	\item Available slit dimensions: $0.5'',1'',1.5'',5''\times12''$ length.
	
	\item Optical efficiency of pre-slit: Averaged over the range $350-1800$ nm, Nasmyth focal plane to slit should be $>80\%$ (ADC included). 

\end{itemize}

\section{Overall description}
The SOXS UV-VIS spectrograph is based on a novel concept in which the incoming beam is partitioned into four polychromatic beams (quasi-orders) using dichroic mirrors, each covering a waveband range of $100-200$ nm. Each quasi-order is diffracted by a custom-made ion-etched grating. The four beams enter a three-element catadioptric camera that images them onto a common detector. The goal of the partitioning is to maximize the overall system throughput.

\subsection{Optical layout}

\begin{table}
	\centering
	\caption{\label{tab:1st-order-params} UV-VIS Spectrograph 1st Order Parameters}
	\begin{tabular}{ll}
Collimator Focal Ratio &	6.5 \\
Collimator Beam diameter &	45 mm \\
Spectral range &	350-850 nm \\
Resolution (1 arcsec slit) &	3500-7000 \\
Slit scale & 110 $\mu \rm{m} / ''$ \\
Foreseen slit widths &	$0.5'' - 1.0'' - 1.5'' - 5''$ \\
Slit height &	$12''$ \\
Camera Output Focal Ratio &	3.11 \\
Detector & Teledyne	e2V CCD44-82 \\
Detector Scale &	$\sim53 \; \mu \rm{m}/''$ \\
Main Disperser &	Four custom ion etched gratings \\
Working temperature &	Ambient $(-5^\circ \rm{C}- +25^\circ \rm{C})$
	\end{tabular}
\end{table}

\begin{figure}
	\centering
	\includegraphics[width=0.65\textwidth]{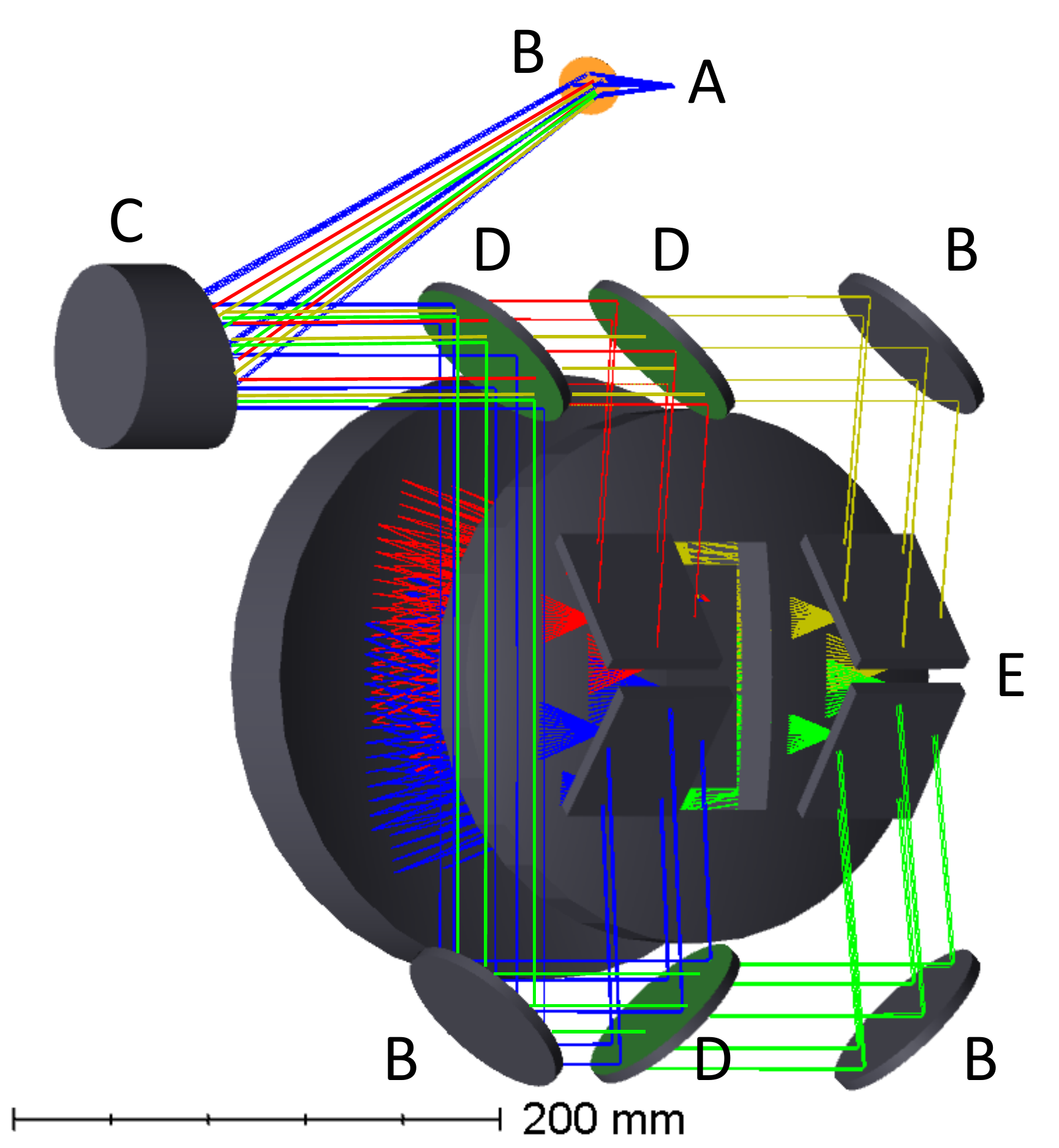}
	\caption{Optical layout of the system. The beam is partitioned by dichroic mirrors and imaged by a single camera. Detailed views of the feed and camera are shown in Figures \ref{fig:mits-feed} and \ref{fig:mits-camera} respectively. A, slit. B, mirror. C, OAP. D, dichroic mirror. E, gratings.\label{fig:mits-layout}}
\end{figure}

\begin{figure}
	\centering
	\includegraphics[width=0.65\textwidth]{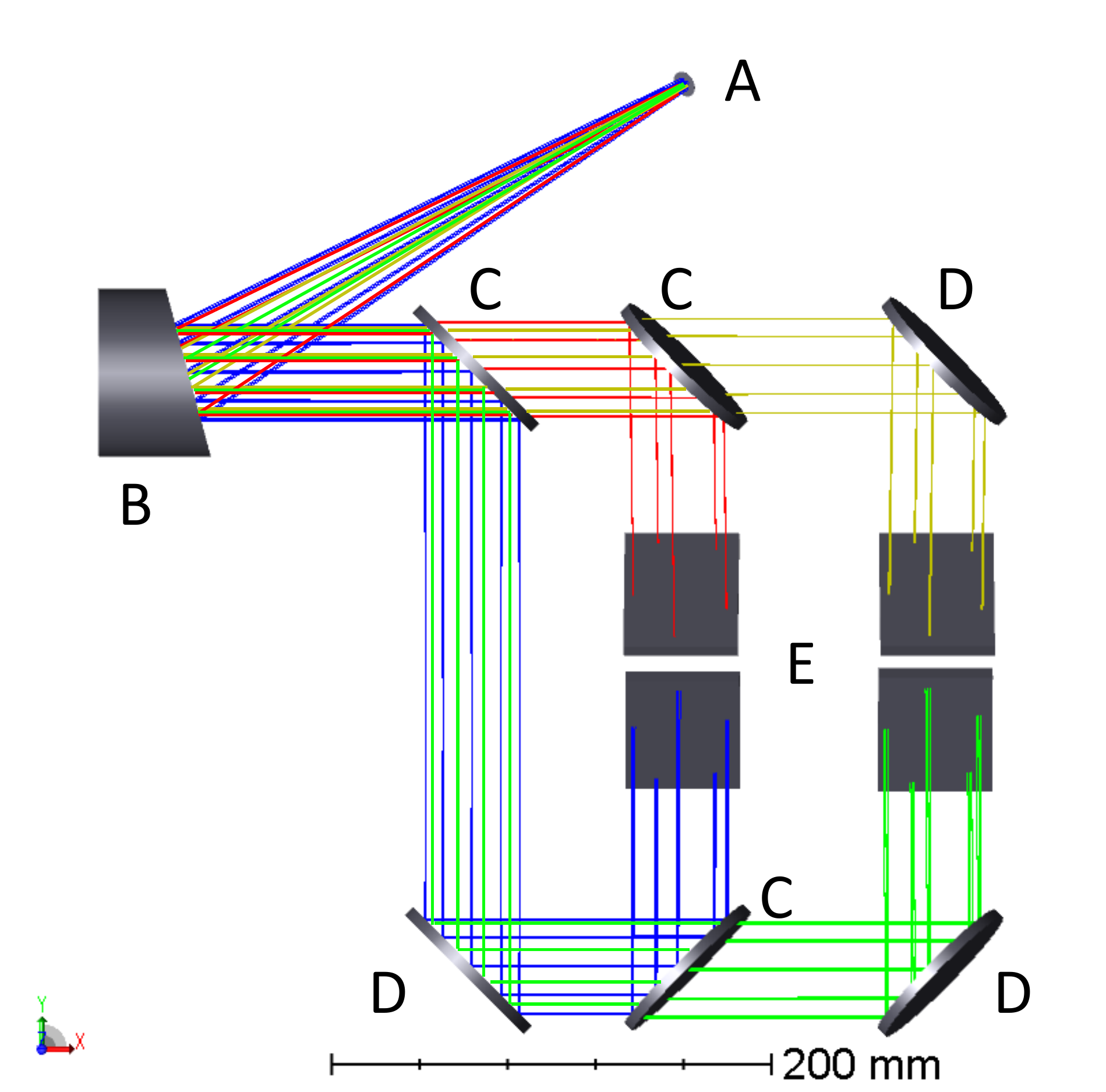}
	\caption{Feed showing the division into four spectral bands. A, fold mirror directing the beam from the slit towards the OAP. B, OAP. C, Dichroic mirror. D, Dielectric mirror. E, Ion-etched gratings. Blue, green, red, and yellow lines correspond to $u, g, r,$ and $i$ configurations respectively. Before each split the rays overlap.\label{fig:mits-feed}}
\end{figure}

The optical layout is shown in Figure \ref{fig:mits-layout}. The first order parameters can be found in Table \ref{tab:1st-order-params}. The beam from the slit is directed by a fold mirror towards an off-axis-parabola (OAP) collimator. It is then divided by a series of dichroics and mirrors towards the gratings. The feed is shown in Figure \ref{fig:mits-feed}.

The system collimator is a 30$^\circ$ OAP with an effective focal length of 272.55 mm and an off-axis distance of 146.3 mm. The F/\# is matched to the SOXS common path (CP) F/6.5 yielding a beam of diameter $\phi=45$ mm. The F/\# and desired beam width dictate a reflected focal length (RFL) of 292.12 mm. We choose a 30$^\circ$ OAP over a higher angle of incidence (AoI) OAP due to improved manufacturability and coating efficiency for low angle OAPs. 

Our system includes three dichroic surfaces to divide the beam into four quasi-orders. Each quasi-order beam is reflected/transmitted by two dichroic surfaces. The first one is a dichroic mirror tilted at $\sim45^\circ$ reflecting $u+g$  and transmitting $r+i$. The two beams are then further divided by secondary dichroic mirrors which are tilted at $\sim45^\circ$ to ensure proper positioning of the beams on the gratings. The transmitted beams ($g+i$) are finally reflected towards their gratings by dielectric mirrors at $\sim45^\circ$. The mirrors are rotated by a few degrees to feed the gratings at the designed $41^\circ$ AoI. 

\subsubsection{Gratings and resolution}
\begin{figure}
	\centering
	\includegraphics[width=0.45\textwidth]{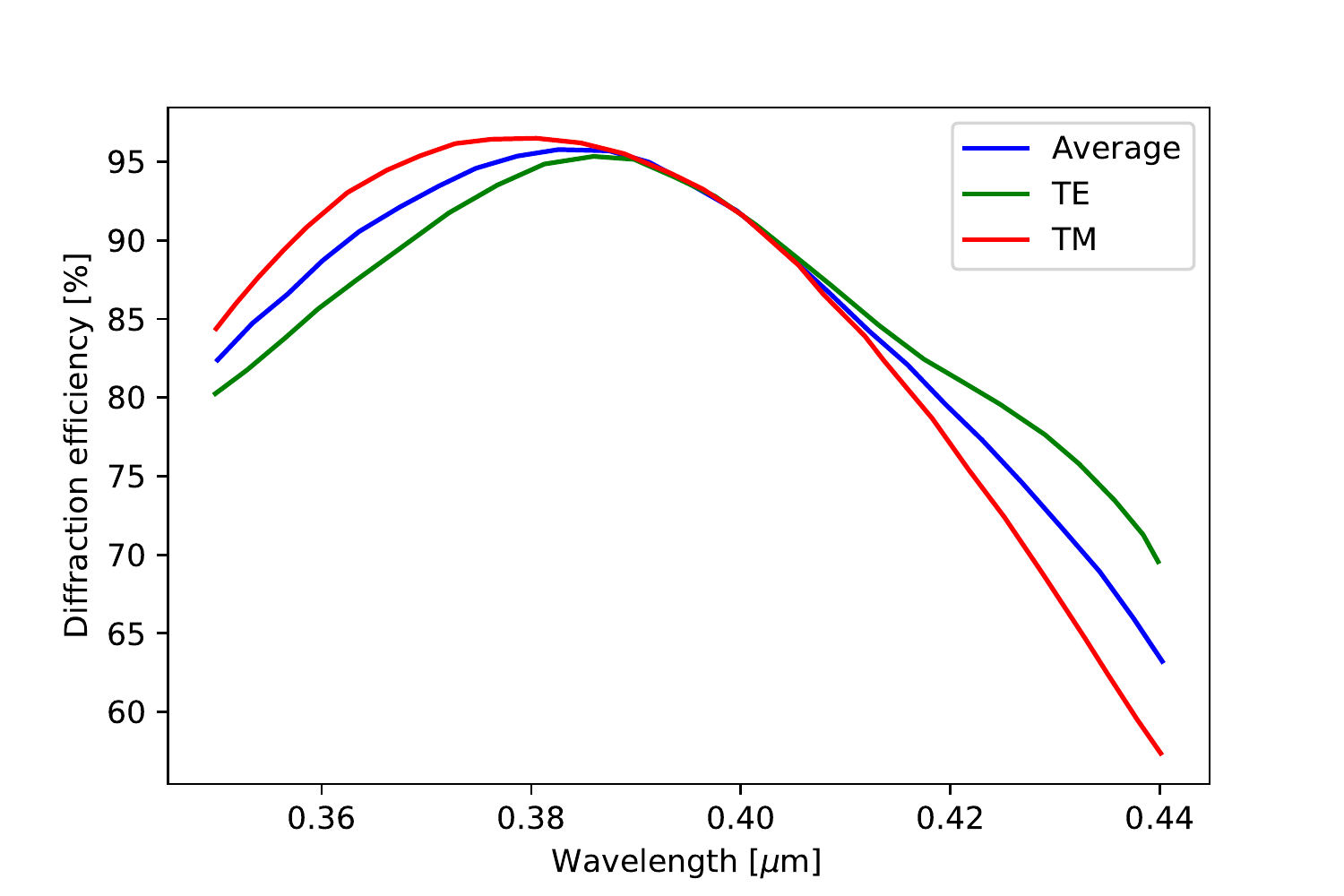}
	\includegraphics[width=0.45\textwidth]{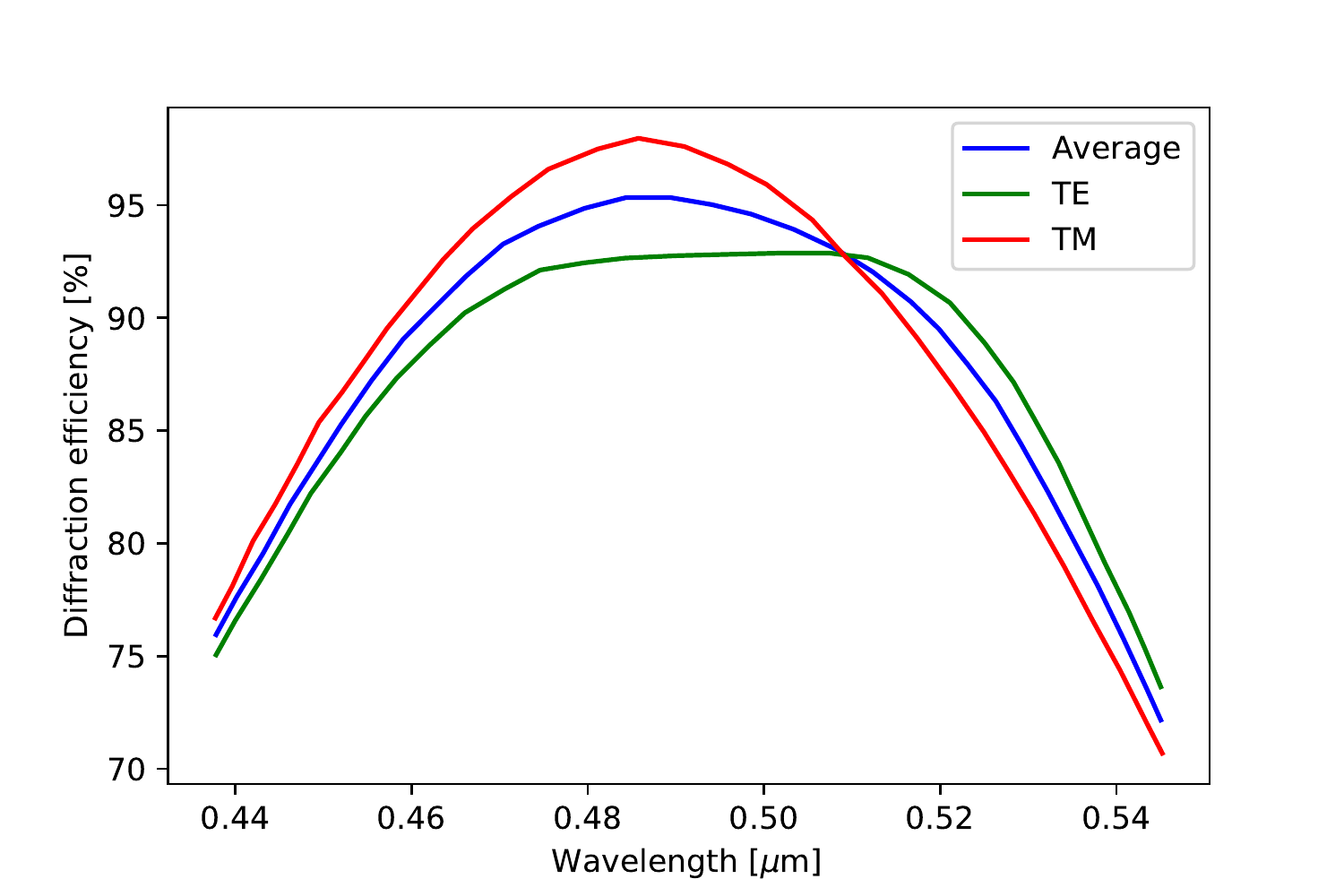}
	\includegraphics[width=0.45\textwidth]{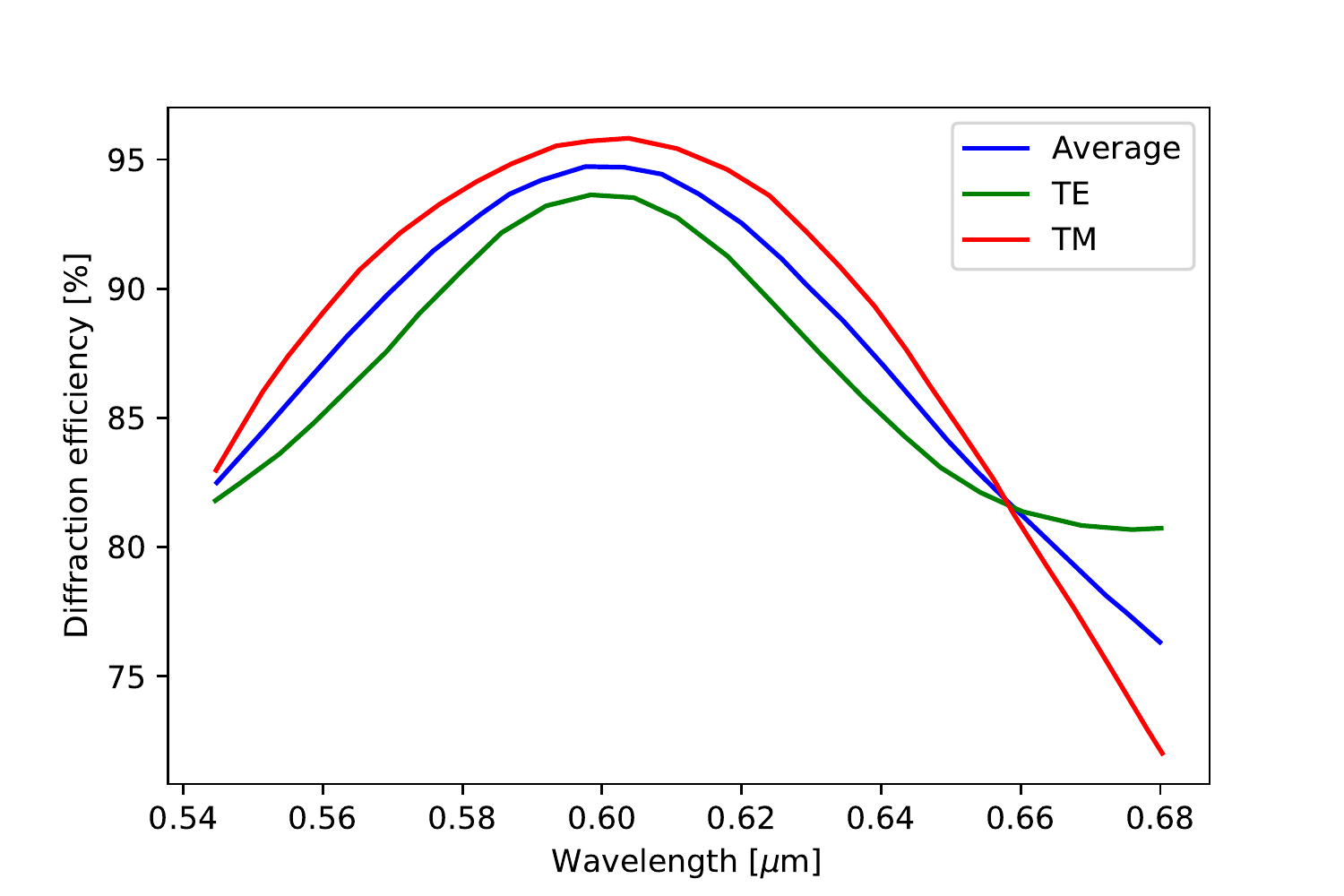}
	\includegraphics[width=0.45\textwidth]{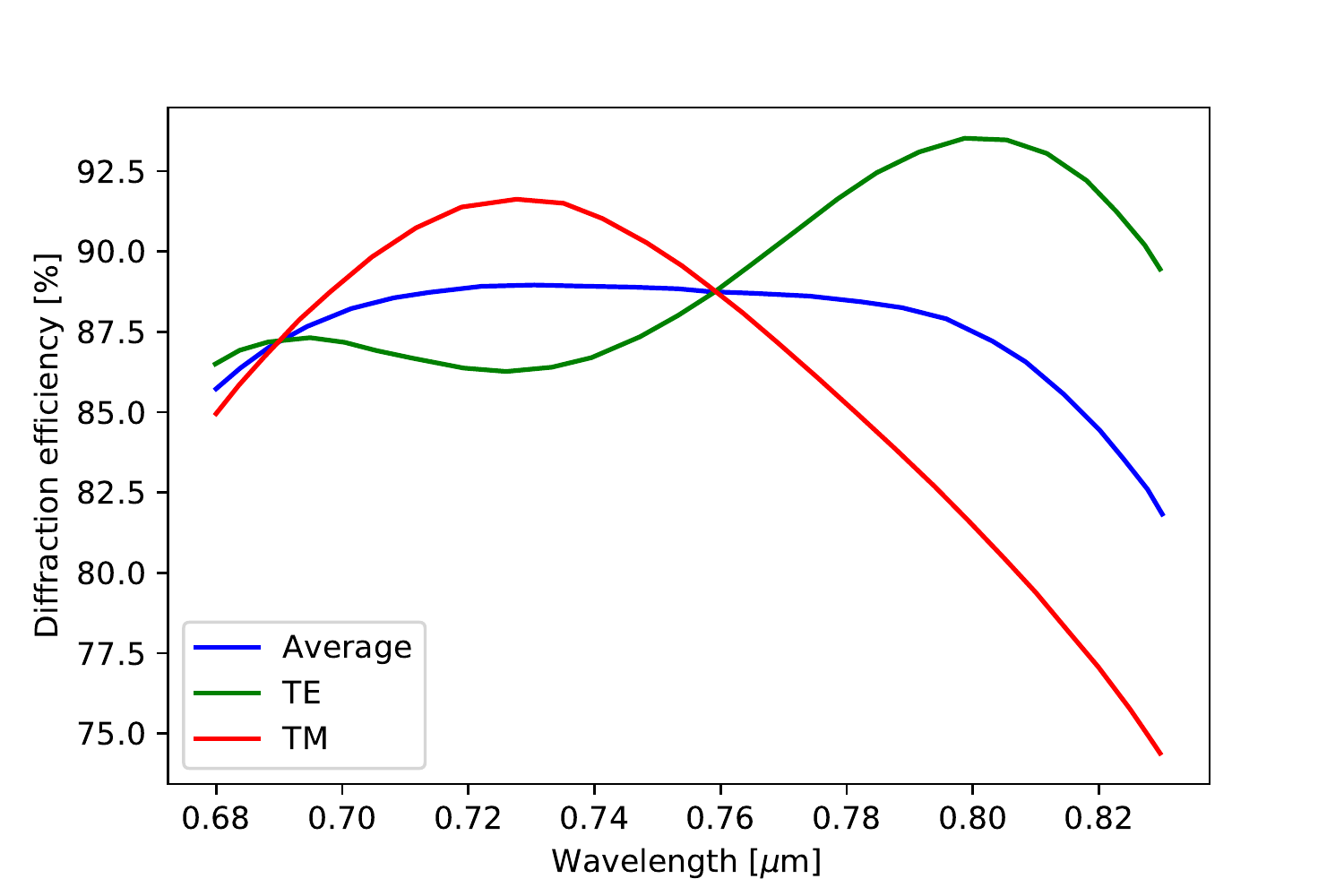}
	\caption{Estimated throughput of ion-etched gratings. Clockwise from top left: $u, g, r,i$. TE and TM represent transverse electric and magnetic polarizations respectively. \label{fig:mits-gratings-throughput}}
\end{figure}

The SOXS UV-VIS arm uses four ion-etched gratings from the Fraunhofer IOF\footnote{Fraunhofer IOF, Albert-Einstein-Str. 7, 07745 Jena.} as the dispersers. We chose the grating parameters (line density, diameter, and feed angle) so that the following conditions are realized:
Littrow angle is roughly $\sim41^\circ$ at each quasi-order center to maximize efficiency.
Resolution for a $1''$ slit is $>3500$  across the entire band, and is $>4000$ for at least 75\% of the quasi-order.
The grating parameters are given in Table \ref{tab:mits_grating_params}. The main advantage of these gratings over \'echelle is that they can be used at first order (m=1), giving an increased efficiency $(\sim80\%)$ across the band. Through discussions with manufacturers we also found these grating to have superior average expected efficiency than similar volume phase holographic (VPH) gratings with the same line densities. The estimated efficiencies of the gratings are given in Figure \ref{fig:mits-gratings-throughput}.

\begin{table}
	\centering
	\caption{SOXS UV-VIS and grating parameters.\label{tab:mits_grating_params}}
	\begin{tabular}{ccccc}
	Quasi-Order & Wavelength Range & Line Density & Feed Angle & $\lambda_{Littrow}$  \\
	& [nm] & [lines/mm] & [$^\circ$] &[nm] \\
	$u$ & $350-440$ & 3380 & $41^\circ$ & 388.2 \\
	$g$ & $427-545$ & 2655 & $41^\circ$ & 494.2 \\
	$r$ & $522-680$ & 2070 & $41^\circ$ & 633.9 \\
	$i$ & $656-850$ & 1660 & $41^\circ$ & 790.4	
	\end{tabular}
\end{table}

\subsection{Camera}
\begin{figure}
\centering
\includegraphics[width=0.7\textwidth]{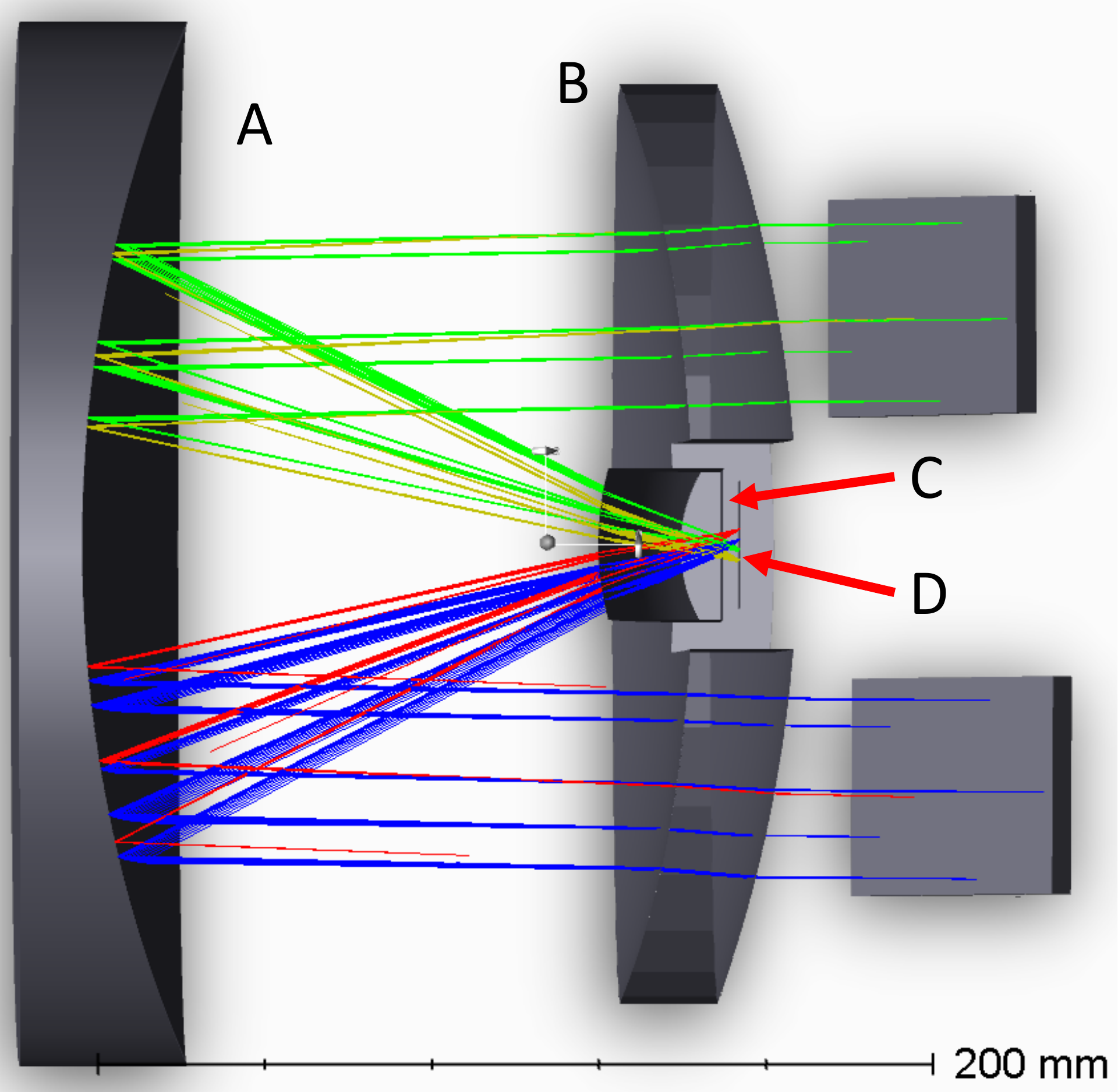}
\caption{Layout of camera. A, Fused silica aspheric mirrror. B, CaF$_2$ aspheric corrector. C, Fused silica field flattener. D, Detector. \label{fig:mits-camera}}
\end{figure}

\begin{figure}
\centering
\includegraphics[width=0.7\textwidth]{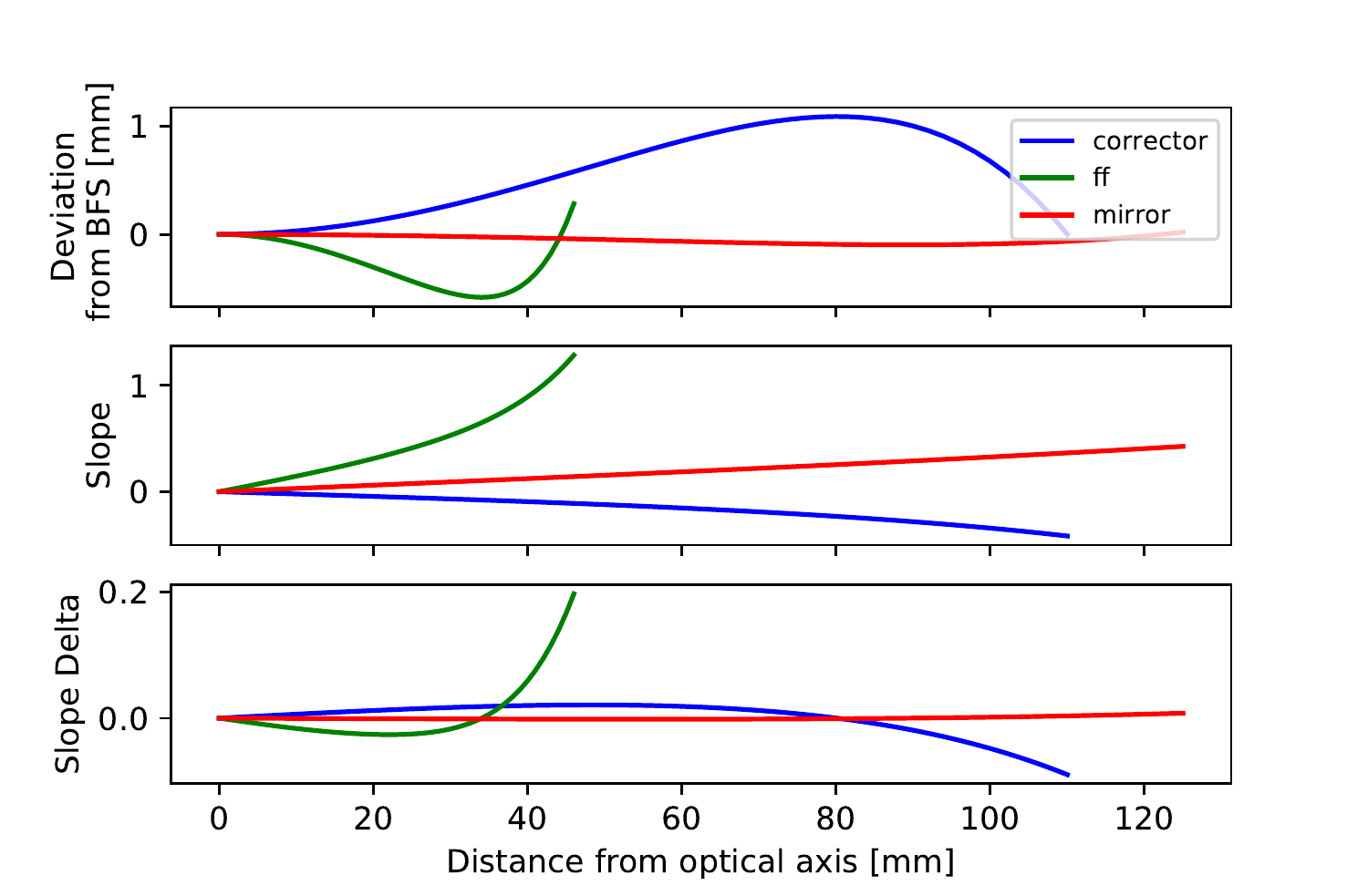}
\caption{Plot of the aspheric sag for the corrector, mirror and field flattener aspheric surfaces.\label{fig:mits-aspheric-sag}}
\end{figure}

The optical prescription of the camera is shown in Table \ref{tab:mits-camera-prescription}, and its layout is shown in Figure \ref{fig:mits-camera}. The camera is made of three aspheric elements and is inspired by the camera designed for MOONS \cite{oliva_toward_2016,delabre_full_2017}. The first element is a CaF$_2$ aspheric corrector with a rectangular aperture at its center. The second element is an aspheric fused silica mirror. The third and last element is an aspheric field flattener which is cut into a rectangular shape, and serves also as the window of the detector cryostat. The sag of the aspheric surfaces is shown in Figure \ref{fig:mits-aspheric-sag}. The aspheric departures are not challenging for modern manufacturing techniques.

\begin{table}
\centering
\caption{Camera prescription. Lens units are in mm.\label{tab:mits-camera-prescription}}
\begin{tabular}{ccccccc}
	 
Name & Radius & Thickness & Material & Clear aperture & Mechanical  & Asphere $4th,6th,8th$ \\
 &  &  &  & Semi-diameter &  Semi-diameter & \\

Aspheric Corrector & 324.88 & 25.00 & CaF$_2$ & 93 & 110 &  \\ 
&&&&&& 2.20E-08 \\
 & 456.46 &  &  & 93 & 110 &	3.87E-13	 \\
 &&&&&& 9.21E-18 \\
Air & Infinity & 146.31 &  &  &  & \\
&&&&&& 9.83E-10 \\
Mirror  & 328.28 &  & MIRROR & 108 & 125.00 & 	5.34E-14	\\
&&&&&&  -7.06E-19 \\
Air & Infinity & 123.61 &  &  &  & \\
&&&&&&  4.81E-07 \\
Detector Window & 70.24 & 30.00 & F SILICA & $15\times42$ & $18\times46$ & 	-6.88E-11	 \\ 
&&&&&& 9.06E-14 \\
 & Infinity & 4.00 &  & $3.5\times35$ & $18\times46$ & \\  
\end{tabular}
\end{table}

\subsubsection{Detector}
\begin{figure}
	\centering
	\includegraphics[width=0.7\textwidth]{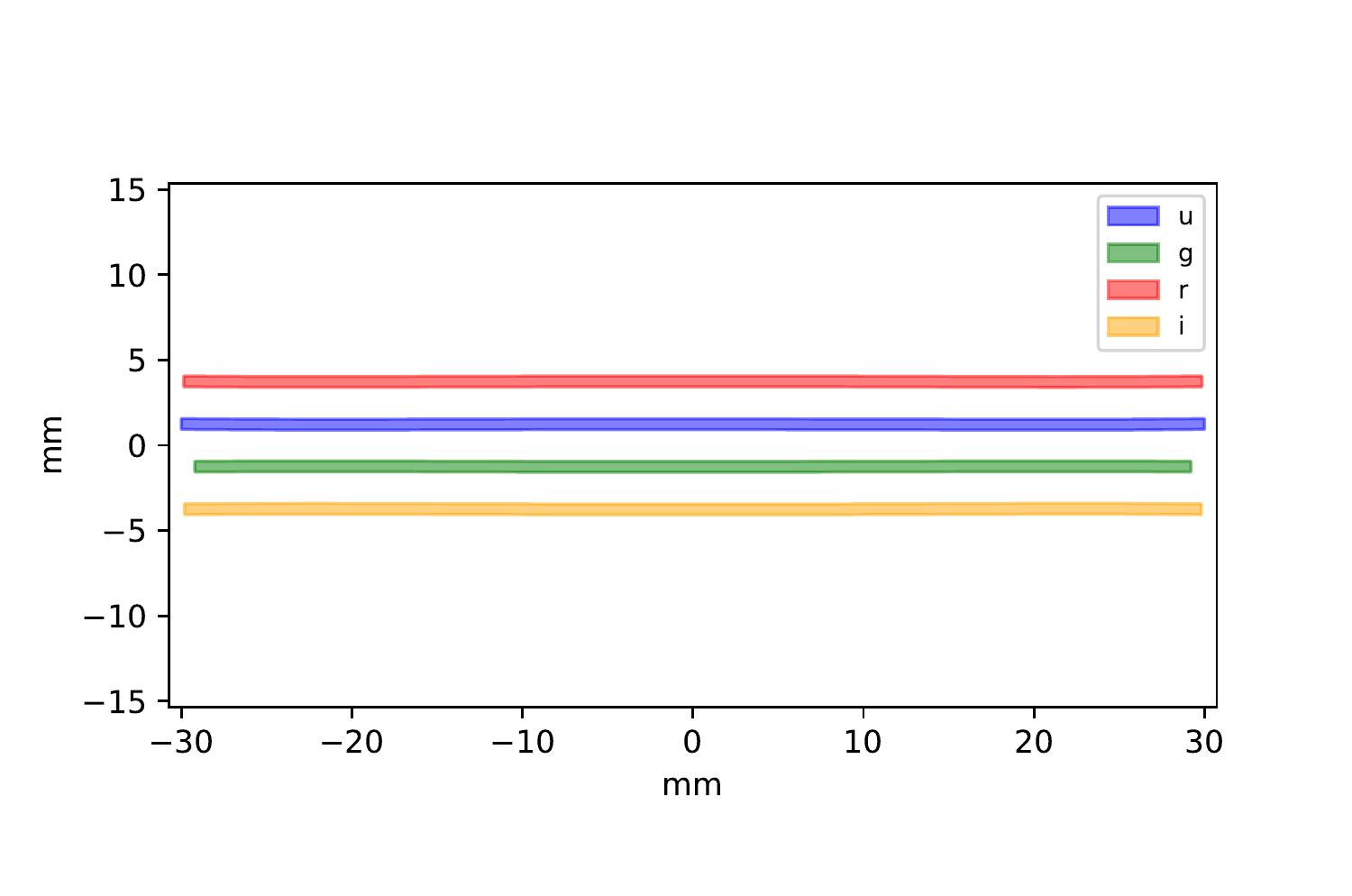}
	\caption{Footprint on the detector including overlap regions. The different colors represent the different quasi orders. The long axis is the spectral direction.\label{fig:mits-detector-format}}
\end{figure}

The detector is a Teledyne e2V CCD44-82 2k$\times$4k back illuminated CCD\cite{soxscosentino} with $15 \; \mu \rm{m} \times 15 \; \mu \rm{m}$ pixels. The spectrum is imaged on the detector in four quasi-orders. The format on the detector is shown in Figure \ref{fig:mits-detector-format}. Note that due to the 41$^\circ$ AoI of the gratings, the spatial axis is slightly rotated relative to the spectral axis. This is due to the need to orient the dispersed beam such that the central wavelength enters the corrector parallel to the optical axis of the camera. The rotation is $\sim8^\circ$ which completes the refracted angle from $2\times41^\circ=82^\circ$ to $90^\circ$. This rotation is small, is typical also of \'echelle spectra, and is easily treated by the data reduction pipeline. The spectral layout on the detector is given in Figure \ref{fig:mits-detector-format}.

\section{Performance}

\begin{figure}
	\centering
	\includegraphics[width=0.7\textwidth]{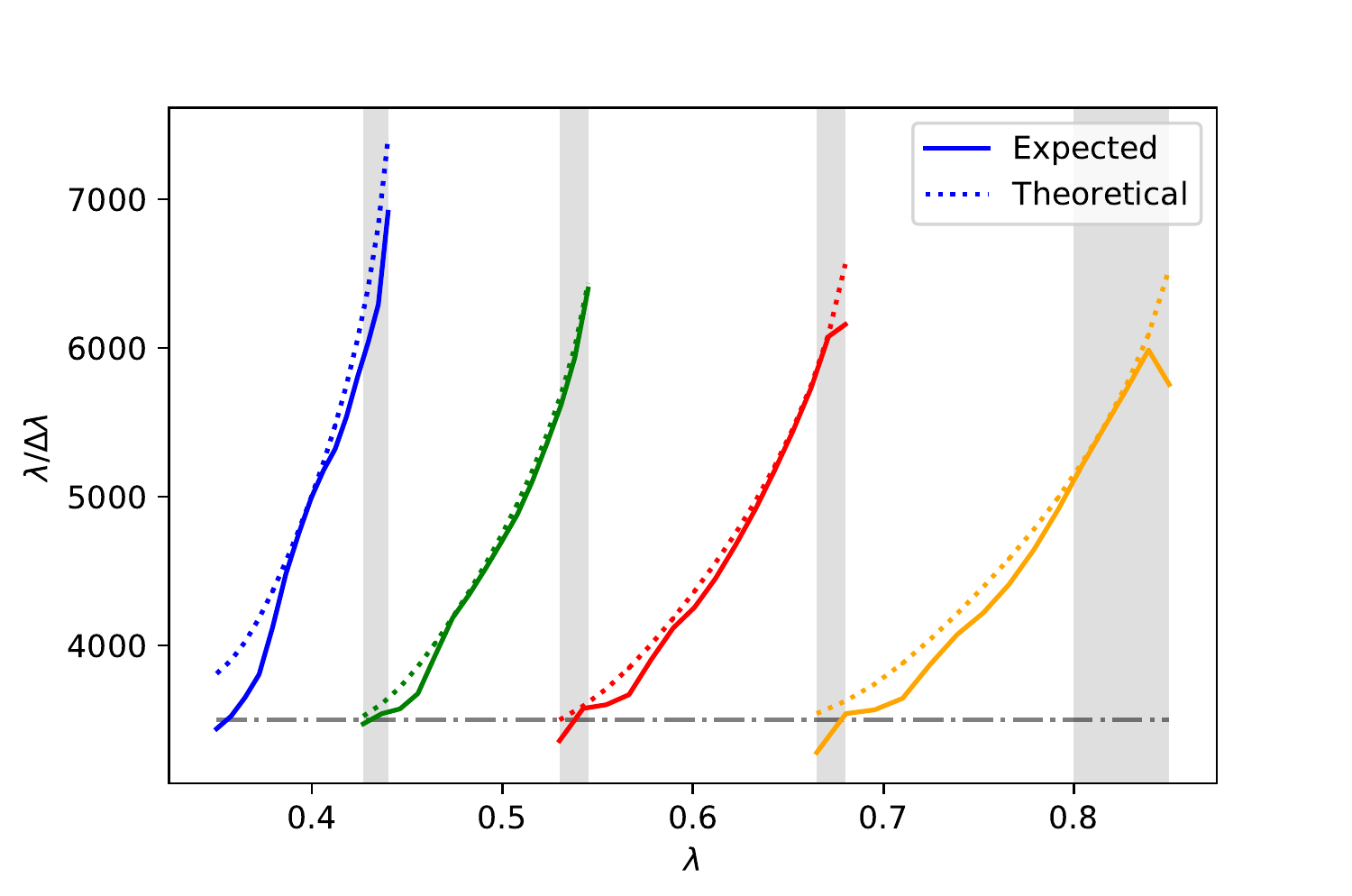}
	\caption{Spectral resolution for $1''$ slit. The shaded regions are the spectral overlap regions both between the quasi orders and between the UV-VIS and NIR arms. The dotted lines are the theoretical resolution plots for the gratings given their characteristics. This figure shows only minimal degradation due to optical aberrations.\label{fig:mits-resolution}}
\end{figure}

\begin{figure}
\centering
\includegraphics[width=0.7\textwidth]{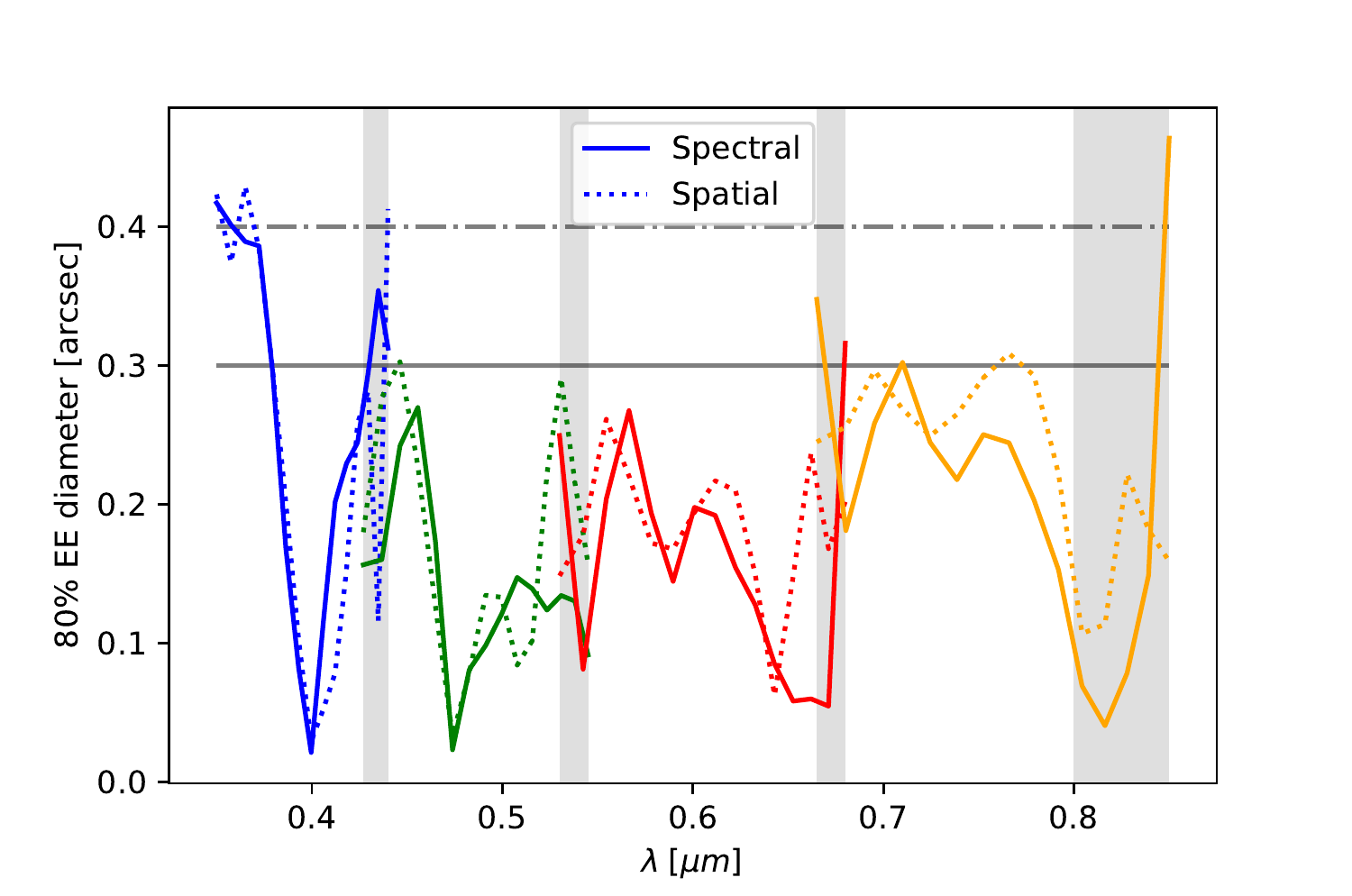}
\caption{Image quality (measured by enslitted energy --- EE) in both spectral and spatial direction. The shaded regions are the spectral overlap regions both between the quasi-orders and between the UV-VIS and NIR arms.\label{fig:mits-image-quality}}
\end{figure}

\begin{figure}
\centering
\includegraphics[width=1\textwidth]{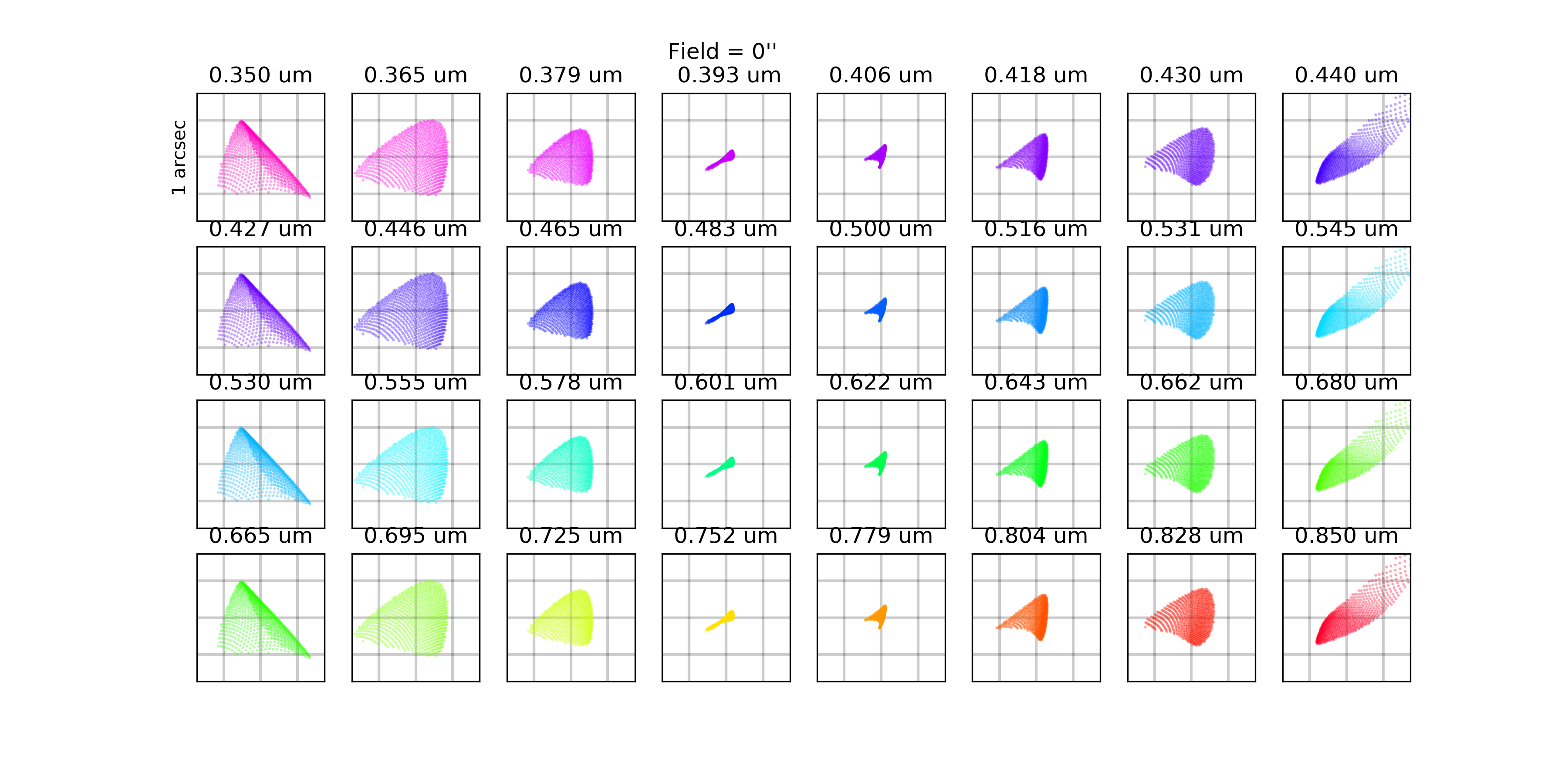}
\caption{PSF for a source at the center of the slit. Each square on the grid is the size of one pixel on the detector.\label{fig:mits-psf-on-axis}}
\end{figure}

The resolution vs. wavelength is reported in Figure \ref{fig:mits-resolution}. We show that the optical design suffers from few aberrations and gives resolution that is close to the theoretical limit. The average resolution and standard deviation across each quasi-order are given in Table \ref{tab:mits-average-resolution}. The resolution is calculated by ray tracing the centroid position on the detector for each wavelength, taking into account the slit width ($1''$), and the diameter of the enslitted energy. The wavelength solution is interpolated with an 8th degree polynomial. The image quality in both the spectral and spatial directions is shown in Figure \ref{fig:mits-image-quality}. The PSF is well contained within $2\times2$ pixels for most of the band and is shown in Figure \ref{fig:mits-psf-on-axis}.

\begin{table}
	\centering
	\caption{Average Resolution.\label{tab:mits-average-resolution}}
	\begin{tabular}{ccc}
		Quasi-Order	& $\overline{R}$ & $\sigma(R)$ \\
		u & 4995 & 1059 \\
		g & 4634 & 890 \\
		r & 4633 & 909 \\
		i & 4601 & 878
	\end{tabular}
\end{table}

\subsection{Throughput}
We assume the following regarding throughput:
\begin{enumerate}
	\item 98.5\% efficiency for broadband ($350-850$ nm) reflective coatings
	\item 99\% efficiency for narrow band ($\Delta \lambda \sim 100$ nm) reflective coatings
	\item 99\% efficiency for AR coatings averaged over the spectral band
	\item Dichroic and grating values are based on data from vendors
\end{enumerate}

With regard to all estimates we take an additional 2\% contingency. Based on these considerations and the quotes we have obtained from manufacturers we expect the throughput of SOXS to be $>60\%$ including contingency. A breakdown of the throughput values is given in Table \ref{tab:mits-throughput}.

\begin{table}
	\centering
	\caption{Expected throughput of the SOXS UV-VIS spectrograph, taking a $2\%$ contingency on all quoted values.\label{tab:mits-throughput}}
	\begin{tabular}{lccccc}
Quasi-order	&		&	$u$	&	$g$	&	$r$	&	$i$	\\
Fold Mirror	&	Expected	&	0.985	&	0.985	&	0.985	&	0.985	\\
&	w/ contingency	&	0.965	&	0.965	&	0.965	&	0.965	\\
OAP	&	Expected	&	0.985	&	0.985	&	0.985	&	0.985	\\
&	w/ contingency	&	0.965	&	0.965	&	0.965	&	0.965	\\
First dichroic	&	Expected	&	0.986	&	0.986	&	0.969	&	0.969	\\
&	w/ contingency	&	0.966	&	0.966	&	0.950	&	0.950	\\
Fold Mirror (u+g)	&	Expected	&	0.990	&	0.990	&	---	&	---	\\
&	w/ contingency	&	0.970	&	0.970	&	---	&	---	\\
Second dichroic (u/g or r/i)	&	Expected	&	0.986	&	0.969	&	0.986	&	0.969	\\
&	w/ contingency	&	0.966	&	0.950	&	0.966	&	0.950	\\
Fold Mirror (g,i)	&	Expected	&	---	&	0.990	&	---	&	0.990	\\
&	w/ contingency	&	---	&	0.970	&	---	&	0.970	\\
Disperser	&	Air-glass interfaces	&	0.990	&	0.990	&	0.990	&	0.990	\\
&	Grating efficiency	&	0.864	&	0.884	&	0.880	&	0.877	\\
&	w/ contingency	&	0.838	&	0.858	&	0.854	&	0.851	\\
Camera	&	Corrector surface 1	&	0.985	&	0.985	&	0.985	&	0.985	\\
&	Corrector surface 2	&	0.985	&	0.985	&	0.985	&	0.985	\\
&	Mirror	&	0.985	&	0.985	&	0.985	&	0.985	\\
&	FF surface 1	&	0.985	&	0.985	&	0.985	&	0.985	\\
&	FF surface 2	&	0.985	&	0.985	&	0.985	&	0.985	\\
&	Total	&	0.927	&	0.927	&	0.927	&	0.927	\\
&	w/ contingency	&	0.909	&	0.909	&	0.909	&	0.909	\\
UV-VIS Spectrograph	&	Total	&	0.741	&	0.737	&	0.749	&	0.726	\\
&	w/ contingency	&	0.643	&	0.627	&	0.663	&	0.630	\\
Common Path	&		&	0.820	&	0.820	&	0.820	&	0.820	\\
Telescope	&		&	0.510	&	0.510	&	0.510	&	0.510	\\
Overall	&	Total	&	0.310	&	0.308	&	0.313	&	0.304	\\
&	w/ contingency	&	0.269	&	0.262	&	0.277	&	0.264	\\
	\end{tabular}
\end{table}

\section{Tolerance}

\subsection{Thermal stability}

Due to the very short distances between the camera elements, there is very low sensitivity to thermal effects. The conceptual mechanical design uses a Invar frame to maintain the distance between the camera lenses, and the detector. We expect a shift of 18 $\mu$m for a temperature change of $\Delta T \sim 25^\circ$C. Therefore the camera does not require an active focus mechanism. Moreover, based on the current mechanical design the spectrum is expected to shift on the detector by 0.1 pixel/$^\circ$C, which is within the requirements.

\subsection{Tolerance}

We take a wedge tolerance of 0.05 mm. For the aspheric surfaces we add a surface decenter of 0.125 mm. The aspheric components are toleranced using a TEZI operand with +/- 0.5 $\mu$m irregularity, which is larger than the manufacturer's capabilities.

We performed 200 Monte Carlo simulations using the expected tolerances, drawn from a uniform distribution to keep a conservative estimate. We allow for the distance between the field flattener and the mirror to be adjusted to achieve the best performance (this can be achieved with shims in the assembled system). The field flattener nominal position is shifted by +/- 135 $\mu$m and is limited to a range of +/- 300 $\mu$m.  The system is most sensitive to tilt and decenter of the camera optics. The design will be re-optimized as the melt data is provided, and again when the as-manufactured optics data will be provided. A full alignment plan is currently under development.

\begin{table}
	\centering
	\caption{Tolerances used. \label{tab:mits-tolerances-used}}
	\begin{tabular}{ll}
Wedge & 0.05 mm \\
Element tilt/decenter & 0.1 deg/mm \\
Index (Abbe) & 0.001 (1\%) \\
Radius & 0.1\% \\
Thickness & 0.05 mm \\
Surface decenter & 0.125 mm \\
Irregularity on aspheres (TEZI) & +/- 0.5 $\mu$m
	\end{tabular}
\end{table}

\begin{figure}
	\centering
	\includegraphics[width=0.7\textwidth]{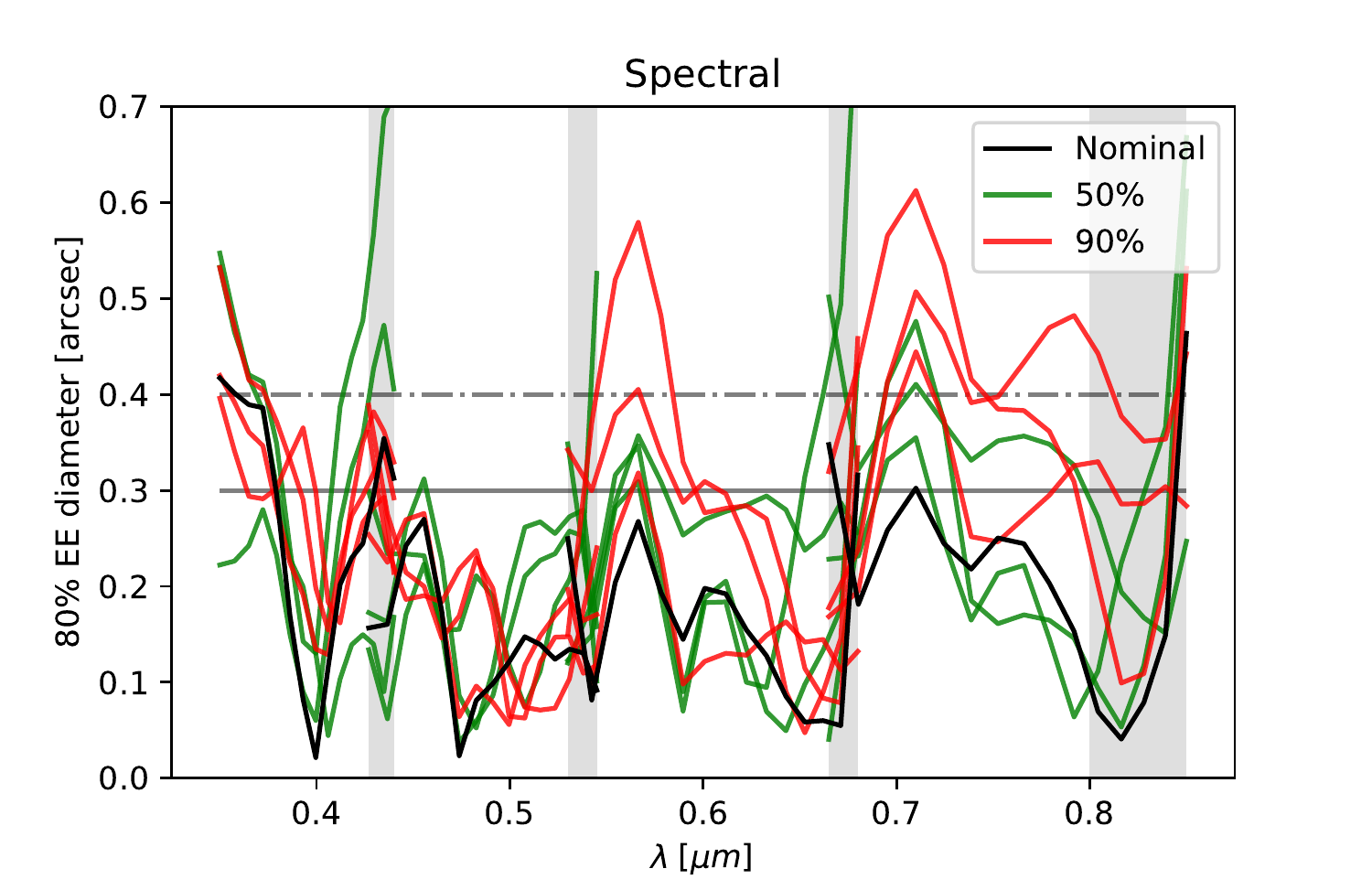}
	\includegraphics[width=0.7\textwidth]{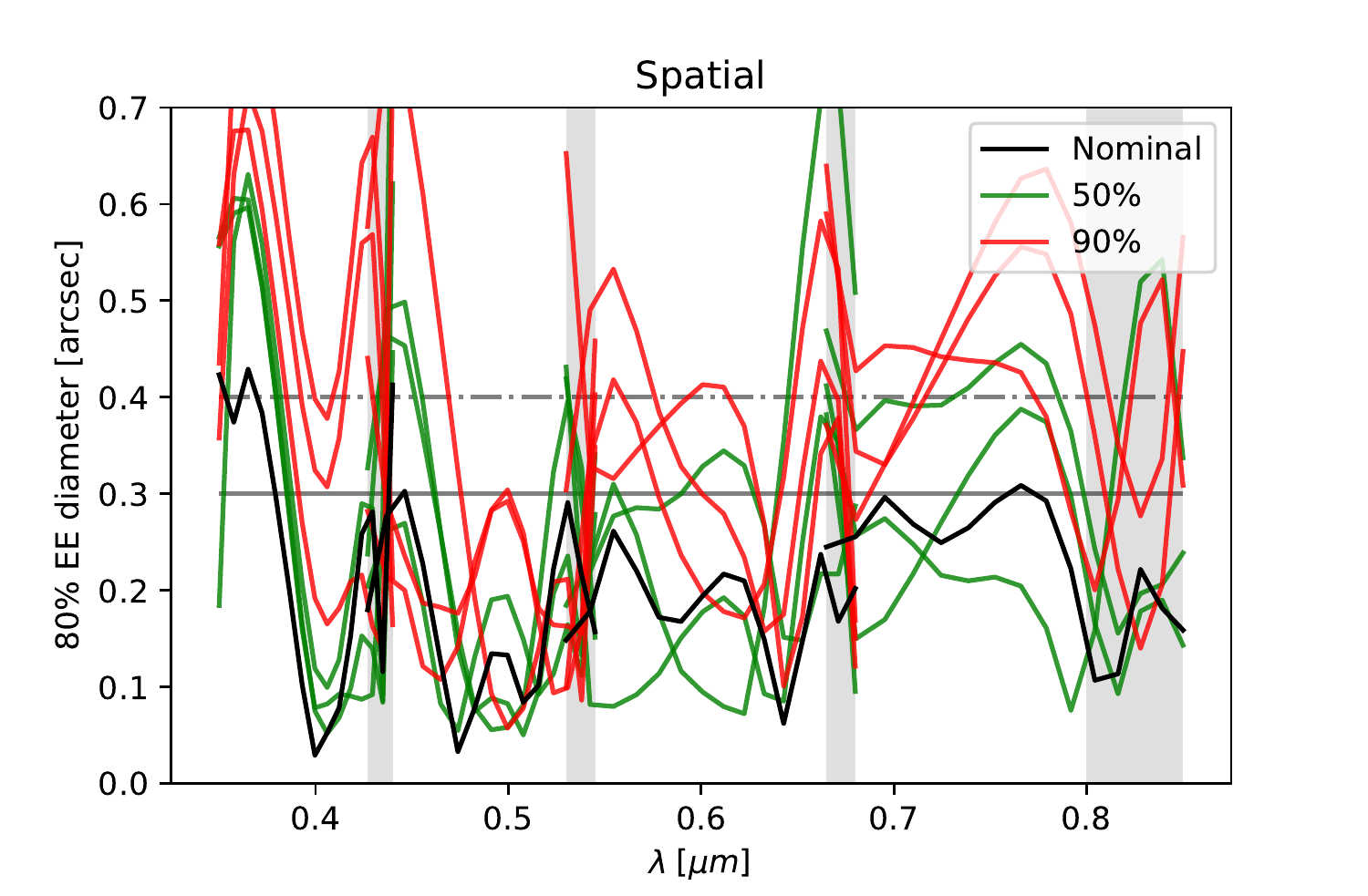}
	\caption{Resulting performance in spectral (spatial) top (bottom) directions. The nominal system performance is shown in black, and the three systems closest to the median and 90th percentiles are shown in green and red respectively. The shaded regions are the overlap between the quasi-orders and the overlap between the UV-VIS and NIR arms of SOXS.\label{fig:mits-tol}}
\end{figure}

\subsection{Ghost analysis}
We identify four potential ghosts that can appear in the system:
\begin{itemize}
	\item Internal reflection in the corrector
	\item Reflection from the planar surface of the field flattener re-reflected by it's front surface
	\item Reflection from the detector which is back reflected by the front of the field flattener
	\item Reflection from the detector which is back reflected by the back of the field flattener
\end{itemize}

\begin{figure}
	\centering
	\includegraphics[width=0.45\textwidth]{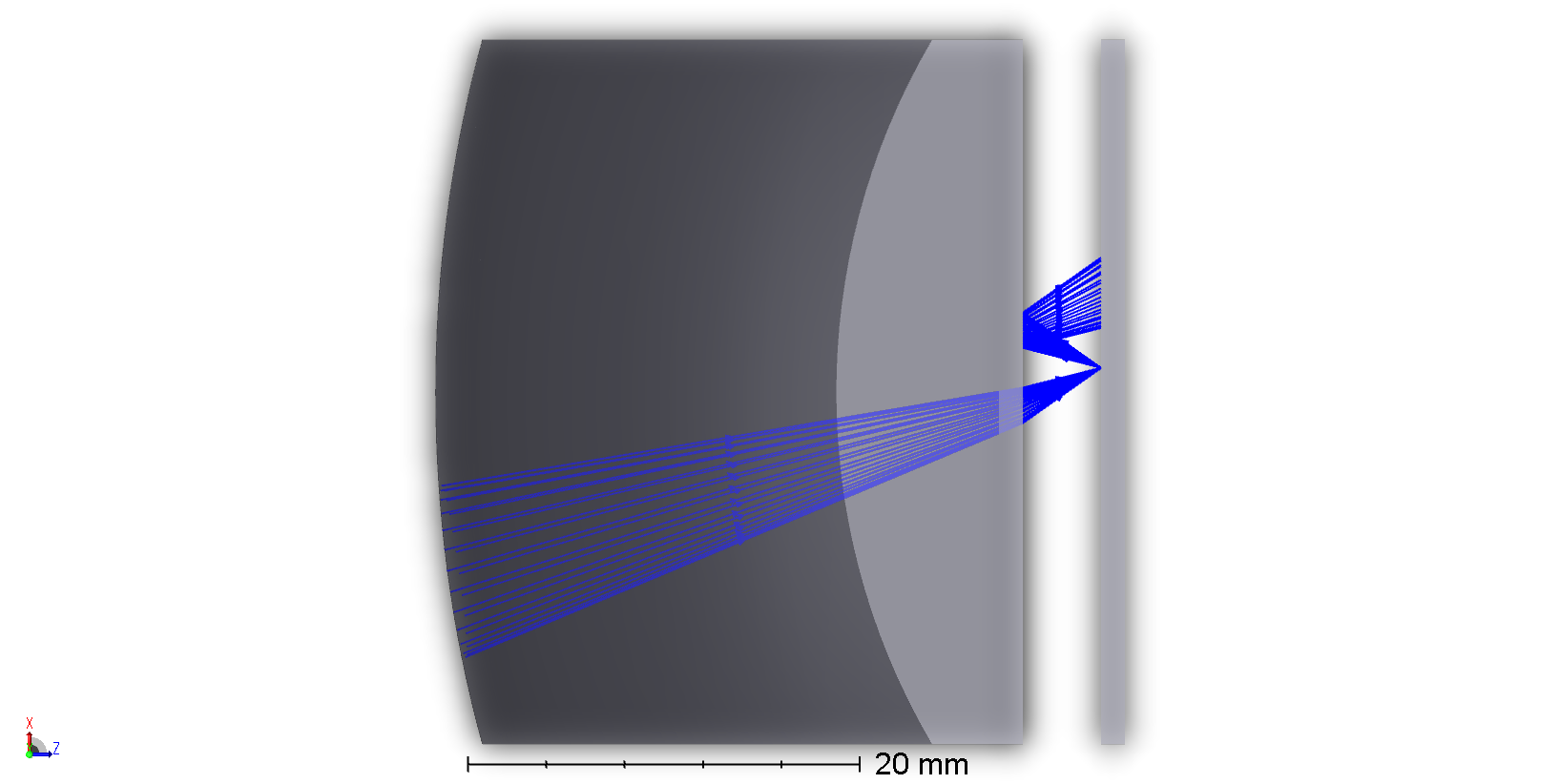}
	\includegraphics[width=0.45\textwidth]{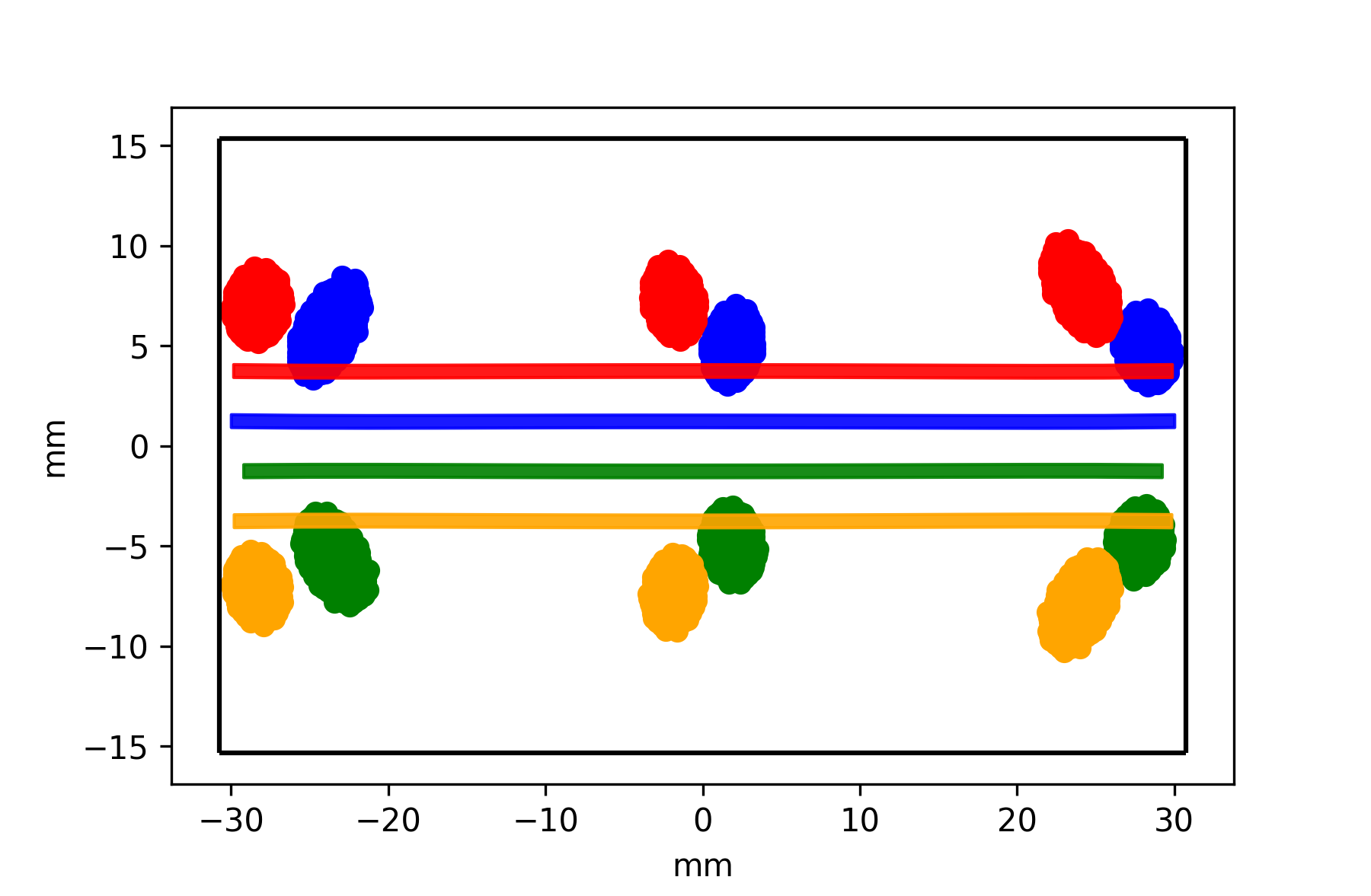}
	\caption{Ghost analysis. Left, end-view of the field flattener showing 400 nm rays being reflection from the back of the field flattener. Right, footprint of the ghost for selected wavelengths (edges and center of band) for each of the quasi-orders.\label{fig:ghost_back_ff}}
\end{figure}

For each of the ghosts we construct a sequential model that simulates the effect and trace rays for three wavelengths (short, central, long) for each quasi-order. Only reflection from the detector and back from the planar surface of the field flattener overlaps with the trace (Figure \ref{fig:ghost_back_ff}). However, this reflection is diffuse with a spot diameter of $\sim 5$ mm; therefore there is a 1/300 attenuation due to defocus, in addition to a 10\% reflection by the detector and a 2\% reflection by the field flattener AR coating. The relative intensity of this ghost is $1/300 \times 0.1 \times 0.02= 6 \times 10^{-6}$. We conclude that this ghost poses no risk.

\subsubsection{Spectral leakage and 2nd order diffraction}

\begin{figure}
	\centering
	\includegraphics[width=0.7\textwidth]{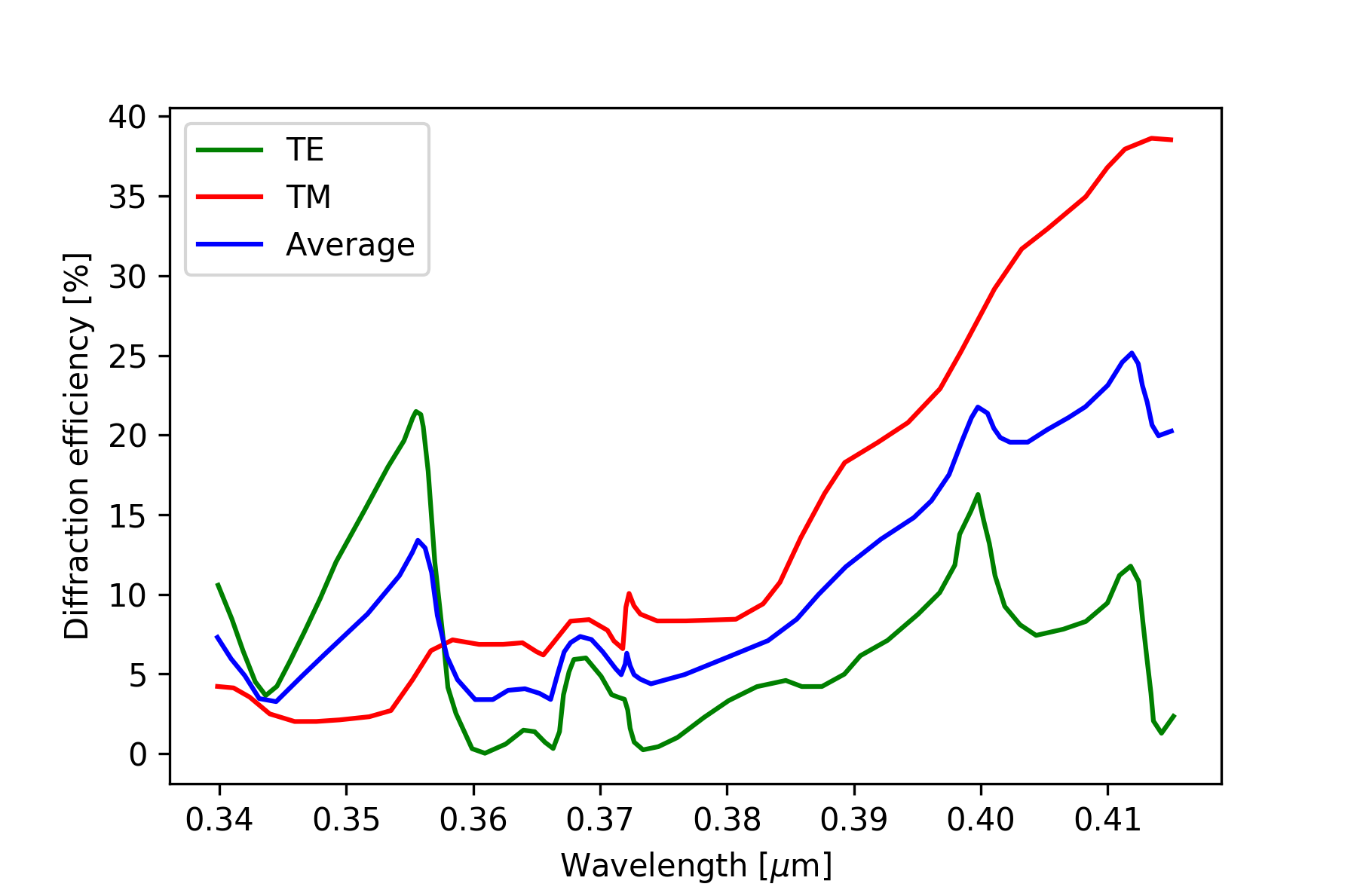}
	\caption{Estimated efficiency of parasitic $m=2$ transmission (private communication with Fraunhofer IOF). TE and TM represent transverse electric and magnetic polarizations respectively.\label{fig:2nd_order_efficiency}}
\end{figure}

Any spectral leakage from the dichroics will arrive at the gratings at the designed AoI. Therefore, for $m=1$ they will be imaged off of the detector and pose no risk. The only residual risk is from u-band light leaking to the $i$-band grating and being diffracted at 2nd order ($m=2$). The i-band grating sits behind two dichroics with 1\% leakage leading to a $0.01^2$ suppression (max $0.05^2$ at specific spikes).
A further suppression of order $10\%-20\%$ would be due to optimization of the grating for 1st, as opposed to 2nd, order. A simulation showing the $i$-band 2nd order efficiency is shown in Figure \ref{fig:2nd_order_efficiency}. In the unlikely event that there is still unacceptable stray light from $m=2$, it is possible to insert an order blocking filter before the $i$-band mirror.

\section{Mechanics}

\begin{figure}
	\centering
	\includegraphics[width=0.7\textwidth]{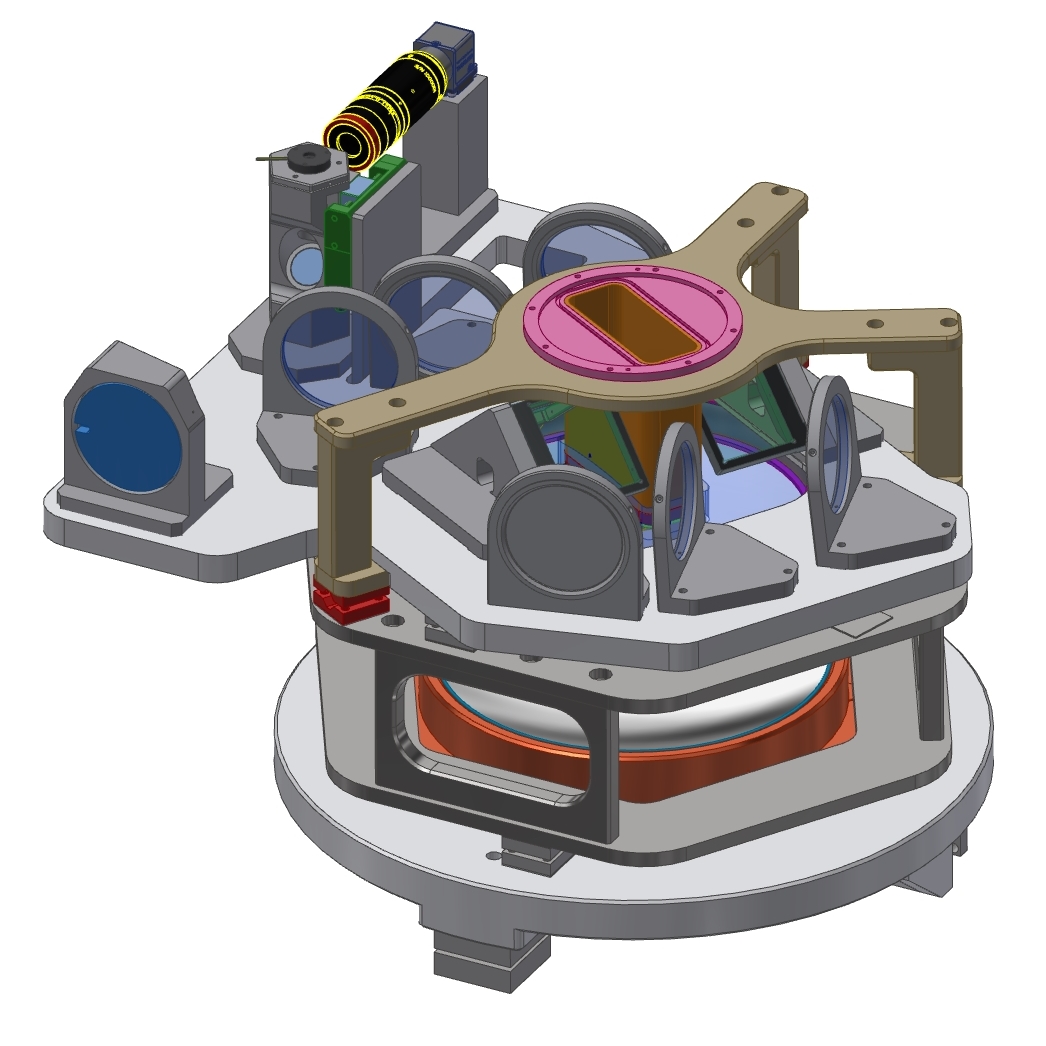}
	\caption{Mechanical design of the UV-VIS arm of SOXS. The camera, dichroics, gratings, OAP, and slit viewing camera are visible.\label{fig:mech}}
\end{figure}

The mechanical structure must support the camera (three elements), the gratings, the dichroics, the OAP, the slit mechanism and slit-viewing camera, and the detector/cryostat assembly. The mechanical system is comprised of four parts which are connected via kinematic mounts: an Al6061 base which connects the UV-VIS system to the NTT flange, an Invar frame which holds the camera primary mirror, a stainless steel frame which holds the CaF$_2$ corrector and field flattener, and an Al6061 plate which holds the feed small optics. The mechanical structure is shown in Figure \ref{fig:mech}. The subsystem is mounted on a sphere-cylinder-flat kinematic mount with the sphere and cylinder positioned along the long axis of the slit to ensure the slit location relative the NIR arm.

\begin{figure}
	\centering
	\includegraphics[width=0.7\textwidth]{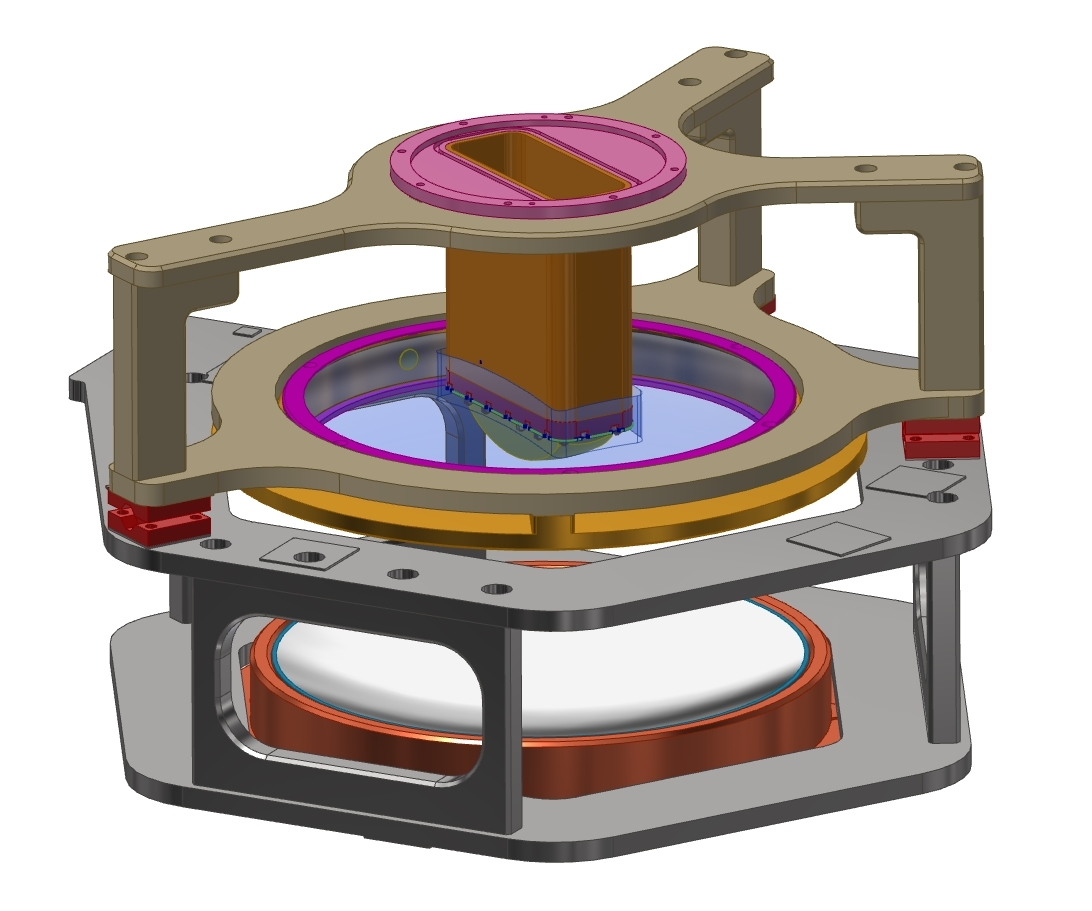}
	\caption{Mechanical design of the camera mount. From bottom: the Invar frame with fused silica primary mirror mounted, then the stainless steel frame with corrector and field flattener mounted.\label{fig:mech-cam}}
\end{figure}

\begin{figure}
	\centering
	\includegraphics[width=0.7\textwidth]{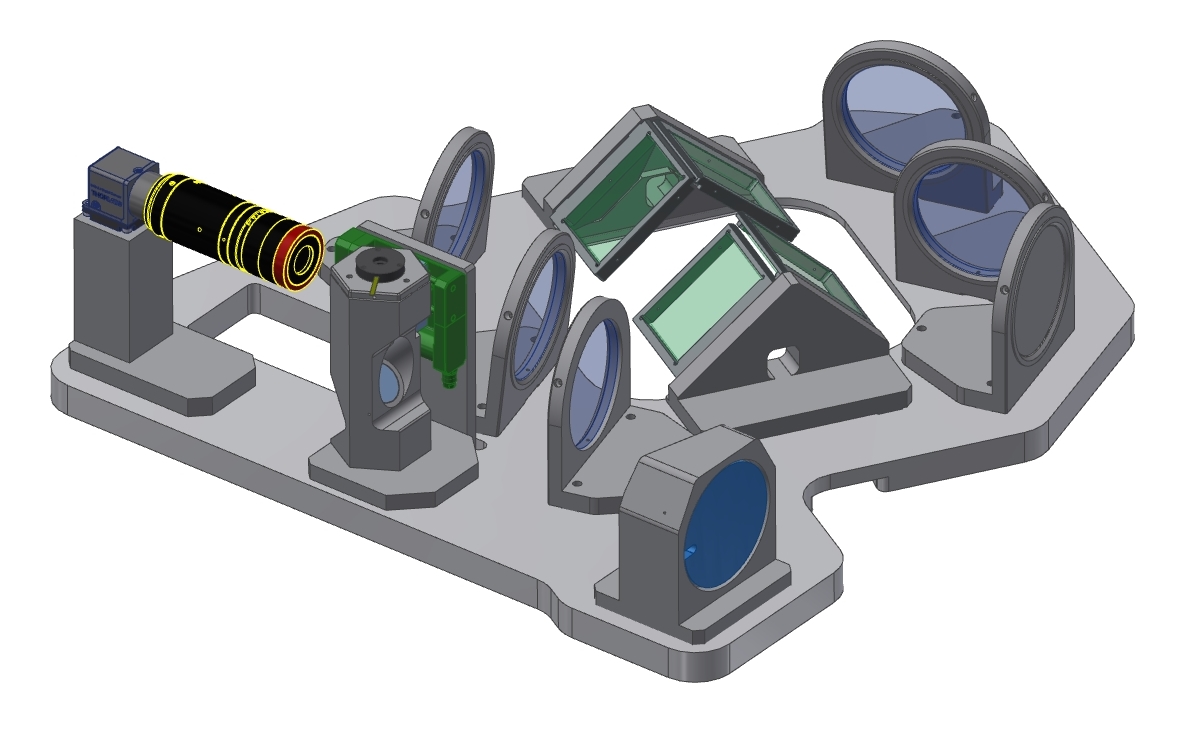}
	\caption{Mechanical design of the feed and small optics mount. The dichroics, gratings (green), OAP, slit mechanism, and slit-viewing camera are visible.\label{fig:mech-feed}}
\end{figure}

The camera mount is shown in Figure \ref{fig:mech-cam}. The stainless steel frame is connected to the Invar frame with a three-cylinder kinematic mount. This mount ensures the camera centration, while the Invar frame ensures the distance between the corrector and the primary mirror is insensitive to thermal changes. The stainless steel frame also counteracts the extension of the stainless steel tube which holds the field flattener. The feed plate (Figure \ref{fig:mech-feed}) holds the Al6061 mounted dielectric and dichroic mirrors, the OAP, and the slit mechanism. The slit mechanism also has a prism position to reflect light towards the static slit-viewing camera.

\section{Summary}

We have shown the design for the SOXS UV-VIS arm. The design makes use of high efficiency ion-etched gratings at used at first order ($m=1$) to provide high throughput---at least 50\% higher than conventional \'echelle medium resolution designs. To accommodate these efficient gratings in the design, the spectrum is divided into four quasi-orders with dichroic mirrors. The dispersed beam is imaged by a single fast camera based conceptually on the camera designed for MOONS. The fast Schmidt catadioptric camera is made of three aspheric elements: a CaF$_2$ corrector, a mirror, and a fused silica field flattener. The design provides spectral resolution $\lambda/\Delta \lambda>3500$ for the whole band, with an average of $\lambda/\Delta \lambda \sim 4700$, and does not require special manufacturing techniques to accommodate the tolerances. Once on the NTT at the beginning of 2021, we expect this instrument to become a workhorse for transient science.

% \clearpage

\bibliographystyle{spiebib}
\bibliography{biblio}

\end{document}